# Multiple Weyl and Double-Weyl Points in an Elastic Chiral Lattice

Yao-Ting Wang[1,2]

[1] *Department of Physics, Imperial College London, London, SW7 2AK, United Kingdom*

[2] *Department of Mathematics, Imperial College London, London, SW7 2AK, United Kingdom*

**We show that Multiple Weyl and double Weyl points arise in a chiral elastic system through stacking many two-dimensional honeycomb mechanical structures. On the distinct $k_z$ plane, the band structures calculated from tight-binding model exhibit the presence of Weyl points at Brillouin vertices and double Weyl Points at Brillouin centre. Based on the tight-binding model, we design a practical chiral mechanical structure which can be fabricated by current 3D printing technology. The numerical calculation illustrates several Weyl and double-Weyl points as expected in our analysis of tight-binding model. To verify the topological feature, topological charges of every degeneracy are calculated. Within these Weyl points, we theoretically prove that the existence of topologically protected surface modes, and their robustness against defects are also demonstrated.**

*Introduction* — In the past decade, Weyl semimetals, which are categorised in solid state crystals and its Fermi energy is located exactly at Weyl points (WPs), have been widely studied as a subject of topological materials in condensed matter physics. WPs are characterised by a nodal degeneracy with linear dispersive cones given by Weyl Hamiltonian $H = \sum_{i,j} k_i v_{ij} \sigma_j$, where $k_i$, $v_i$, and $\sigma_i$ are wavevectors, velocities, and Pauli matrices respectively. Intriguingly, instead of opening a gap for the degeneracy, the violation of spatial symmetry merely leads to a shift of WPs in momentum space. This robustness against geometric perturbation exists due to the fact that all elements in Weyl Hamiltonian are used. Furthermore, to determine whether a band crossing is a WP, it is necessary to calculate Berry curvatures and topological orders in the vicinity of the degeneracy. Besides, by mimicking the case of electric field radiated from a charge, Berry curvatures can be regarded as certain fields emitted by a "topological charge" (TC) [1]. Since every WP has nonzero TCs, the distribution pattern of Berry curvature exhibits a source/sink feature around WPs; hence, the volume integration enclosed the corresponding WP gives rise to the value of TCs.

After a substantial number of theoretical and numerical predictions of the Weyl semimetals [2-10], several experimental observations of Weyl quasiparticles were proposed in electronic [11-15] and bosonic regimes [16-20]. Apart from that, a new type of topological semimetals named "Type-II Weyl semimetals" were proposed to differ from the usual (Type-I) Weyl semimetal. Due to the violation of Lorentz symmetry, a type-II Weyl semimetal has tilted Weyl cones which make Fermi energy level intersect not only the WPs but also bulk bands [21]; it leads to an unusual density of states compared to Type-I WPs. Experimental observations of type-II Weyl fermions were discovered in a layered transition-metal dichalcogenides [22-24] and a single crystalline compound [25-26]. For WPs, in addition to the linearly degenerate signature in all three dimensions, there are topological surface modes due to non-vanishing topological charges. While topological surface modes connecting WPs were firstly measured in electronic systems [12-13], similar phenomena have been demonstrated extensively in acoustic systems [19], photonic [20], and plasmonic [27] systems.

Aside from the WPs, several theoretical works indicate that so-called double Weyl points (DWPs), which possess higher topological orders, emerge in the solid-state crystals accompanied by specific symmetries. Unlike the single Weyl points (SWPs), which is named to differ from DWPs, the DWP carries TCs whose values are equal to ±2 because DWPs are the coalescence of two SWPs. Generally, in the DWP case, the band crossing is expressed as the quadratic dispersion around the degeneracy in at least one momentum plane, which implies quadratic terms dominates the low-energy Hamiltonian. The observation of DWPs has been proposed in the crystalline solid of $SiSr_2$ [28] and photonic crystals [6,17].

However, although a vast number of researches related to WPs and non-trivial surface modes have been reported, there is still a lack of the elastic counterpart owing to the complexity of elastic solids. In this paper, firstly a tight-binding(TB) model is discussed to pave the way for the following studies in a real elastic system. With the aid of this TB model, we investigate the generation of multiple SWPs and DWPs in an elastic structure consisting of honeycomb chiral lattice. As the DWPs are guaranteed by certain rotational symmetries, breaking those symmetries leads to the separation of the DWP, and then it splits into two SWPs. Also, the Chern number calculation and the exhibition of topologically nontrivial surface states provide another solid evidence to the emergence of WPs and DWPs. As the structure was designed under the fabricating limit of current 3D printer technology, the experimental realisation of such structure is expected in the near future.

*Tight-Binding Model* — We begin by introducing the TB model for an AA-stacked honeycomb lattice. The practical elastic chiral structure will be discussed in the later section. The displacements between two nearest neighbours (NN) in the same layer is $a$ and the lattice constant along z direction is $c$. The translation vectors are expressed as $\mathbf{r}_{lmn} = l\mathbf{a}_1 + m\mathbf{a}_2 + n\mathbf{a}_3$, where the primitive lattice constants $\mathbf{a}_1 = \sqrt{3}a\hat{\mathbf{x}}$, $\mathbf{a}_2 = (\sqrt{3}\hat{\mathbf{x}} + \hat{\mathbf{y}})a/2$, and $\mathbf{a}_3 = c\hat{\mathbf{z}}$. Besides, three longitudinal and transverse unit vectors connecting between the nearest neighbours are given by

$$\hat{\mathbf{L}}_1 = (\sqrt{3}\hat{\mathbf{x}} + \hat{\mathbf{y}})/2, \quad \hat{\mathbf{L}}_2 = (-\sqrt{3}\hat{\mathbf{x}} + \hat{\mathbf{y}})/2, \quad \hat{\mathbf{L}}_3 = -\hat{\mathbf{y}},$$

$$\hat{\mathbf{T}}_1 = (-\hat{\mathbf{x}} + \sqrt{3}\hat{\mathbf{y}})/2, \quad \hat{\mathbf{T}}_2 = -(\hat{\mathbf{x}} + \sqrt{3}\hat{\mathbf{y}})/2, \quad \hat{\mathbf{T}}_3 = \hat{\mathbf{x}}.$$

Consisting of the two p-orbital components $p_x$ and $p_y$, the NN TB Hamiltonian is given by $H = \sum_{lmn} H_{\perp}^L + H_{\perp}^T + H_{\square} + H'$, where

$$H_{\perp}^L = t_{\perp}^L \left[ \left(\mathbf{a}_{lmn}^\dagger \cdot \hat{\mathbf{L}}_1\right)\left(\hat{\mathbf{L}}_1 \cdot \mathbf{b}_{lmn}\right) + \left(\mathbf{a}_{lmn}^\dagger \cdot \hat{\mathbf{L}}_2\right)\left(\hat{\mathbf{L}}_2 \cdot \mathbf{b}_{l-1,mn}\right) + \left(\mathbf{a}_{lmn}^\dagger \cdot \hat{\mathbf{L}}_3\right)\left(\hat{\mathbf{L}}_3 \cdot \mathbf{b}_{l,m-1,n}\right) + \text{H.c} \right], (1)$$

$$H_{\perp}^T = t_{\perp}^T \left[ \left(\mathbf{a}_{lmn}^\dagger \cdot \hat{\mathbf{T}}_1\right)\left(\hat{\mathbf{T}}_1 \cdot \mathbf{b}_{lmn}\right) + \left(\mathbf{a}_{lmn}^\dagger \cdot \hat{\mathbf{T}}_2\right)\left(\hat{\mathbf{T}}_2 \cdot \mathbf{b}_{l-1,m,n}\right) + \left(\mathbf{a}_{lmn}^\dagger \cdot \hat{\mathbf{T}}_3\right)\left(\hat{\mathbf{T}}_3 \cdot \mathbf{b}_{l,m-1,n}\right) + \text{H.c} \right], (2)$$

$$H_\Box = t_\Box \left[ \mathbf{a}^\dagger_{lmn} \cdot \mathbf{a}^\dagger_{lm,n+1} + \mathbf{a}^\dagger_{lmn} \cdot \mathbf{a}^\dagger_{lm,n-1} + \mathbf{b}^\dagger_{lmn} \cdot \mathbf{b}^\dagger_{lm,n+1} + \mathbf{b}^\dagger_{lmn} \cdot \mathbf{b}^\dagger_{lm,n-1} \right], \quad (3)$$

$$H' = \lambda \hat{\mathbf{z}} \cdot \left[ \mathbf{a}^\dagger_{lmn} \times \mathbf{a}^\dagger_{lm,n+1} + \mathbf{a}^\dagger_{lmn} \times \mathbf{a}^\dagger_{lm,n-1} + \mathbf{b}^\dagger_{lmn} \times \mathbf{b}^\dagger_{lm,n+1} + \mathbf{b}^\dagger_{lmn} \times \mathbf{b}^\dagger_{lm,n-1} \right]. \quad (4)$$

where $\mathbf{c}^\dagger$ ($\mathbf{c}$), $\mathbf{c} = \mathbf{a}, \mathbf{b}$ represent the creation (annihilation) vector operator and $\mathbf{a}(\mathbf{b}) = [a_x(b_x), a_y(b_y)]^T$. $t_\perp$ and $t_\Box$ are the hopping parameters for the intralayer and interlayer. $\lambda$ stands for the extra interlayer interaction resulted from the presence of cross-coupling for $p_x$ and $p_y$. After being transformed to the momentum space, the Hamiltonian in terms of wavevectors $\mathbf{k}$ reads

$$H(\mathbf{k}) = \begin{bmatrix} 2t_\Box \cos(x_3) & -i2\lambda \sin(x_3) & t^L_\perp F(x_1,x_2) & t^T_\perp F_+(x_1,x_2) \\ i2\lambda \sin(x_3) & 2t_\Box \cos(x_3) & t^T_\perp F_-(x_1,x_2) & t^L_\perp F(x_1,x_2) \\ t^L_\perp F^*(x_1,x_2) & t^T_\perp F^*_-(x_1,x_2) & 2t_\Box \cos(x_3) & -i2\lambda \sin(x_3) \\ t^T_\perp F^*_+(x_1,x_2) & t^L_\perp F^*(x_1,x_2) & i2\lambda \sin(x_3) & 2t_\Box \cos(x_3) \end{bmatrix}, \quad (5)$$

where $x_j = \mathbf{k} \cdot \mathbf{a}_j$ and the complex elements $F$ and $F_\pm$ are defined by $F = 1 + \sum_{j=1}^{2} \exp(-ix_j)$, $F_\pm = 1 + \sum_{j=1}^{2} \exp[-i(x_j \pm 2\pi j / 3)]$. With Eq. (5), we are in the position of discussing SWPs and DWPs via analysing the above TB model. From Fig. 1a to 1c, we depict the band structures along distinct reduced Brillouin boundaries in accordance with the hopping parameters $t^L_\perp = 1$, $t^T_\perp = -0.1$, $t_\Box = 0.1$, and $\lambda = 0.3$. Evidently, while $k_z = 0$ and $\pi/c$, there exist two quadratic degeneracies at $\Gamma$ point between 1st - 2nd and 3rd - 4th bands, and one linear degeneracies at K point between 2nd - 3rd bands. It is worth noting that, for $k_z \neq 0$ or $\pm\pi/c$, this fixed wave vector in z direction generates effectively synthetic gauge fields [5] which make the system behave like a topological Chern insulator in $k_x$-$k_y$ sub-reciprocal plane.

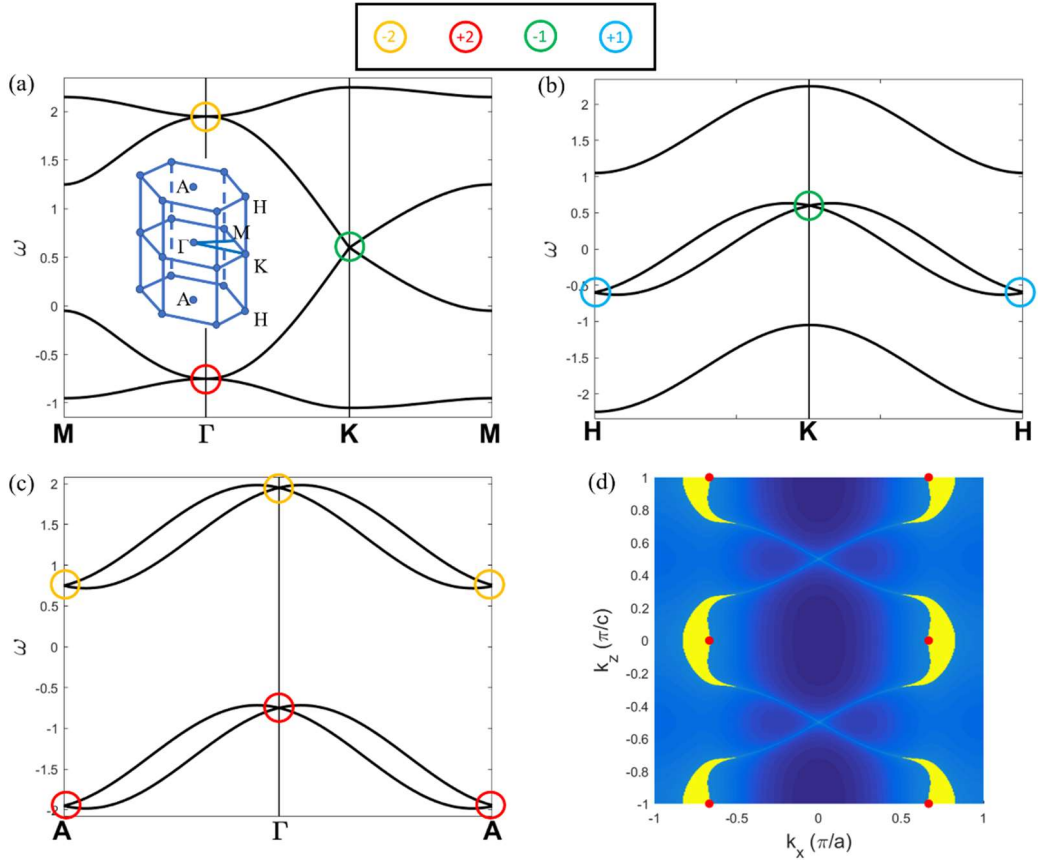

Fig. 1(a) The band structure based on Eq. (5). One SWP and two DWPs with distinct topological charges are labelled in colour. The inset illustrates the reciprocal hexagonal lattice of the TB model. Band structures along the (b) HKH and (c) AΓA boundary. Especially, at Γ and A point, the line shapes of dispersions are quadratic in $k_x$-$k_y$ reciprocal plane but remain linear as $k_z$ changes. (d) Equi-frequency contour at zero frequency. The yellow areas are bulk regions and the red dots mark the location of SWPs. Within WPs, Fermi arcs are clearly seen.

To clarify the topological characteristics of all these band crossings in this TB model, the TC distribution corresponding to degeneracies needs to be checked. In the classical electromagnetism, Gauss' law bridges the electric field distribution and charge magnitude. In the same manner, provided that the Berry curvature distributions of certain degeneracies behave as a field source or sink, they will carry a non-vanishing TC whose magnitudes can be obtained by taking the volume integral enclosed each degenerate point. From the Fig. 1a to 1c, the values of TCs are labelled in different colours to display WP(DWP)s possessing ±1(2) charges at high-symmetry points K(Γ) and H(A). In addition, connecting with a positive TC and a negative one, Fermi arcs emerge in the equi-frequency contour as a boundary is cut. For this TB model, we pick up the zero-frequency plane and truncate a surface in *x* axis. As shown in Fig. 1d, while there exist several bulk regions due to the frequency difference between two WPs, Fermi arcs are clearly shown and connect the area around SWPs.

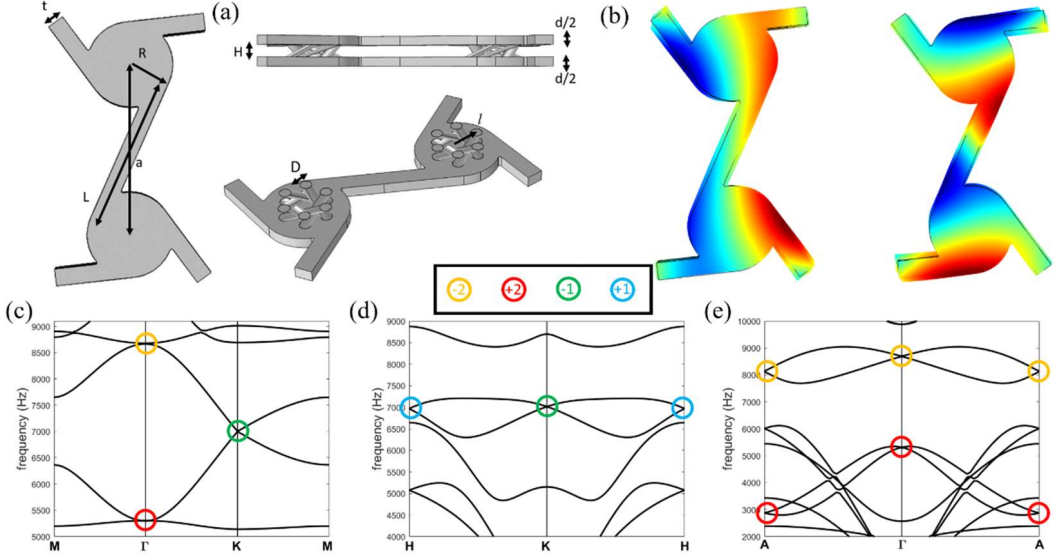

Fig. 2(a) Unit cell of the elastic chiral lattice consisting of polymers. The in-plane chirality annihilates all the mirror or rotational symmetries except $C_6$ symmetry, and the out-of-plane one is in charge of interlayered LT coupling. The geometry parameters are denoted in [29]. (b) The flexural $p_x$-$p_y$ eigenmodes between 1st and 2nd bands at Γ point, where red and blue parts respectively illustrate the positive and negative flexural waves. The band structures from (c) to (e) show similar signatures compared with Fig. 1. Colour circles highlight the positions and magnitudes of WPs in band structures.

*Elastic Chiral Lattice* — The previous TB model discussion has offered the insight as how to design a real elastic structure. According to the parameters given in [29], Fig. 2a illustrates an elastic chiral structure with honeycomb lattice. The structure is made of polymers ($\rho$ = 1190 kg/m³, E = 2.72×10⁹ Pa, $\nu$ = 0.38) which is feasible for recent 3D printing technology [19,30]. The transverse (*x*-*y*) chiral connecting beams eliminate all relevant types of rotational and mirror symmetries except $C_6$ symmetry, by which the isolated degeneracies can be guaranteed. Along the z direction, twisted elastic beams linked with two layers give rise to the flexural $p_x$-$p_y$ mode coupling which forms the cross terms given in Eq. (4). Therefore, the emergence of WPs, DWPs, and topologically non-trivial elastic surface modes in terms of distinct wave vector $k_z$ are intuitively expected. Fig. 2b depicts the flexural dipole eigenmodes, of which this four-band diagram are made. In Fig. 2c-e, we numerically calculate the corresponding band structure via the commercial software COMSOL v5.3. With the proper hopping parameters, the result given by TB model depicted in Fig. 1 possesses similar characteristics compared with the numerical calculation for the bands composed of $p_x$-$p_y$ modes, which shows the previous analysis applies to this elastic chiral system. In Fig. 2d, we present band structure along KH and, together with Fig. 2c, the dispersions in the vicinity of the degeneracies show the linearity in all momentum directions at the frequency around 7000 Hz, which imply they are the candidates of SWPs. In addition to the SWPs, as shown in Fig. 2c, there are two DWPs intersected by $1^{st} - 2^{nd}$ and $3^{rd} - 4^{th}$ bands at Brillouin zone centre. Near these nodes, the main feature of DWPs, whose dispersions are quadratic in transverse **k** plane but linear along $k_z$ direction, is also plotted in Fig. 2c-d.

To further prove the existence of non-vanishing TCs near the degenerate points, we numerically calculate a closed surface integration of Berry curvature enclosing these degeneracies [1,6]. The colour circles from Fig. 2c-e label the TC of different WPs with ±1 and ±2 at Brillouin zone vertices and the centre, which are consistent with the results given by the TB method. In Fig. 2d, due to the chiral connecting beams, breaking all the mirror symmetries causes the SWPs on the same $k_z$ plane to possess the identical charges. Moreover, since the sum of all WPs within the BZ must be neutralised, there must exist two distinct $k_z$ planes having SWPs with opposite TCs, such that $k_z = 0, \pm\pi/c$ in Fig. 2d.

In addition to the direct calculation of topological charges, the TC of a WP can be analytically obtained by the low-energy expression by means of employing symmetry arguments [8,17]. It is well known that $C_3$ symmetry generates Dirac points possessing TC of ±1 at Brillouin vertices K and K'. As a result, we neglect the discussion of this linear degeneracy but focus on the emergence of deterministic DWPs. In the system with $C_3$ rotational and time-reversal invariances, in the following content we theoretically show that, together with both symmetries, they guarantee topologically non-trivial DWPs at certain points [17]. Around these symmetries invariant points, such as Γ and A, the low-energy 2×2 Hamiltonian can be expressed as $H_{eff}(\mathbf{q}) = f(\mathbf{q})\sigma_+ + f^*(\mathbf{q})\sigma_- + g(\mathbf{q})\sigma_z$, where $\mathbf{q}$ represents in-plane small-quantity momentum expansions from the symmetry invariant points. $\sigma_\pm = \sigma_x + i\sigma_y$ are circular superpositions of Pauli $x$ and $y$ components and $\sigma_z$ are Pauli matrix $z$ components. As the basis of above effective Hamiltonian are $(1,0)^T$ and $(0,1)^T$, the matrix representation of $C_3$ symmetry operators is denoted as $\mathbf{C}_3 = \exp(\pm i2\pi\sigma_z/3)$. Since the effective Hamiltonian is invariant under $C_3$ rotation, one obtains

$$e^{\mp i4\pi/3} f(q_+, q_-) = f(e^{i2\pi/3}q_+, e^{-i2\pi/3}q_-), \tag{6a}$$

$$g(q_+, q_-) = g(e^{i2\pi/3}q_+, e^{-i2\pi/3}q_-). \tag{6b}$$

where $q_\pm = q_x \pm iq_y$. Besides, as the time-reversal invariance leads to the relation that $\mathbf{T}H_{eff}(\mathbf{q})\mathbf{T}^{-1} = H_{eff}(-\mathbf{q})$, coefficient function $f$ must be under the time-reversal operation. If $f$ can be expanded as $\sum_{n_1 n_2} A_{n_1,n_2} q_+^{n_1} q_-^{n_2}$, combing the above arguments ensures all linear terms must vanish. It makes the lowest expansions become quadratic, and then TCs of ±2 arise. It is worth noting that, if time-reversal symmetry is broken, DWPs still survive as long as the $C_6$ symmetry in the system is preserved [8].

When the symmetry is broken by adding perturbations, one of the effects for WPs is that, the perturbation only moves the position of WPs in momentum space instead of lifting a gap. As a result, if one breaks the symmetry which guarantees the deterministic DWPs, they will split into two SWPs in the momentum space but make no gaps emerge. In Fig. 3a, the $C_3$ symmetry in the elastic chiral lattice is broken by filling two tungsten carbide plates ($\rho$ = 15520 kg/m³, E = 6×10¹¹ Pa, $v$ = 0.2) in two opposite isosceles triangular shapes into AB sublattice. The side and bottom lengths of triangles equals R/2 and R/3, respectively. Fig. 3b-c illustrates 3D band structures in the $k_z$ = 0 plane that exhibits two SWPs evolving from DWP. Two SWPs are located at the coordinates ($k_x$, $k_y$, $f$) = ($\mp$0.05π/a, $\mp$0.05π/a, 5060 Hz) between 1$^{st}$ and 2$^{nd}$ bands, and ($k_x$, $k_y$, $f$) = ($\pm$0.02π/a, $\mp$0.08π/a, 8204 Hz) between 3$^{rd}$ and 4$^{th}$ bands. The coordinates of SWPs on the same plane are opposite to each other owing to the presence of $C_2$ rotation in $z$ axis. As both SWPs are separated from a DWP, they have the identical charge magnitudes due to the conservation of TCs. The above discussion shows the robustness of WPs against the symmetry violation since the they shift rather than open a gap in the k space.

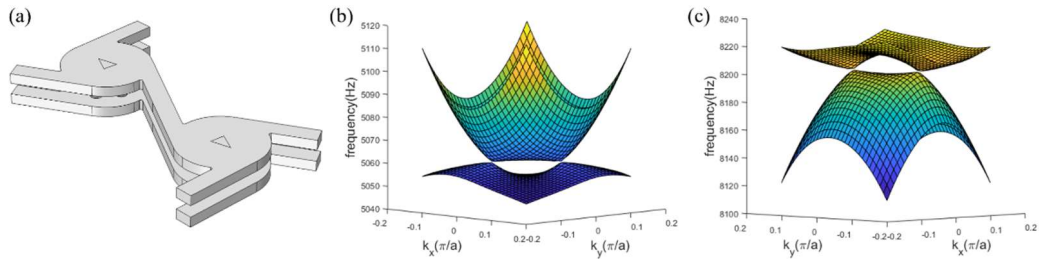

Fig. 3(a) Two oppositely directional isosceles triangles are embedded in the chiral lattice to break $C_3$ symmetry. After eliminating $C_3$ symmetry, two SWPs split from the DWP between (b) 1$^{st}$ – 2$^{nd}$ and (c) 3$^{rd}$ -4$^{th}$ bands arise.

*Robust Surface Modes* — The emergence of topological boundary modes may be one of the most essential properties in topological materials. With a fixed $k_z$ value, the off-diagonal terms resulting from the chiral interlayer coupling in Eq. (5) can be considered as a synthetic gauge field in a 2D subsystem [5-6], and it lifts non-trivial bandgaps where topologically protected boundary modes arise. Fig. 4a presents the band structure with fixed $k_z$ = π/2c. Due to the presence of the band Chern numbers marked in Fig. 4a, there are two topologically non-trivial bandgaps in accordance with the bulk-boundary correspondence. In Fig. 4b, we illustrate the projected band structures on the $k_z$ = π/2c plane, which shows topological surface modes obtained by numerically calculating a 1×15 super-cell elastic chiral lattice with truncation along $x$ axis. The surface modes coloured in red and green represent the upper and lower boundaries, respectively. At $k_z$ = π/2c and $f$ = 6700 Hz, the elastic waves travel unidirectionally along the truncated surface and propagate robustly against the defect, as shown in Fig. 4c. While $k_z$ = ‒π/2c, the elastic wave in Fig. 4b propagates along counter-clockwise direction due to the presence of negative group velocity. Yet, the elastic wave is still under topological protection as it passes through a sharp corner without reflections.

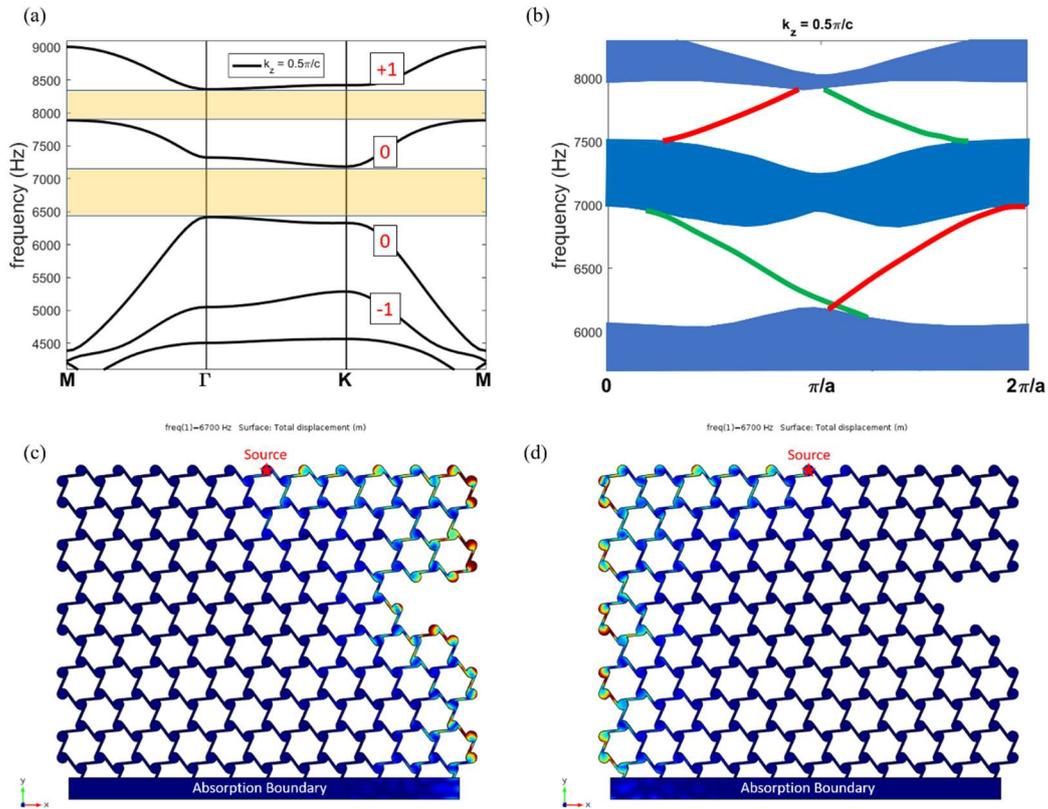

Fig. 4 (a) Fixing $k_z$ values as the synthetic gauge field for the band structure in 2D subdomain. Numbers attached to bands denote the corresponding Chern numbers. (b) Projected band structure with truncation along $x$ axis. Topological surface modes are coloured in red and green to differ from the surface modes located at the upper and lower boundaries. On the different $k_z$ planes, the surface modes with topological protection are demonstrated in (c) and (d).

*Conclusion* — In conclusion, multiple SWPs and DWPs in an elastic chiral lattice have been studied. We begin by a TB model whose band structure exhibits SWPs at Brillouin vertices and DWP at Brillouin centres. The TCs magnitude are verified to further prove the topological characteristics for WPs. Between these TCs, there exist Fermi arcs which have been demonstrated in an equi-frequency contour. To show the validity of this TB model, we have proposed an elastic structure by periodically connecting chiral honeycomb lattices with rotationally linking beams which generate the synthetic gauge fields in a 2D subsystem. The appearances of both band structures are similar because the proper parameters in the TB model are chosen. Additionally, at $k_z = 0$ or $\pm\pi/a$, SWPs and DWPs arise in the same position as given by the TB model. In the final part, we verify the topological boundary modes in the system while $k_z = \pm\pi/2a$. Both cases exhibit topological protection which makes no backscattering as the waves hit a defect.